\documentclass[epj]{webofc}
\usepackage[varg]{txfonts}   
\woctitle{XLVI International Symposium on Multiparticle Dynamics}
\begin{document}
\title{Oscillations in counting statistics}
%
%

\author{Grzegorz Wilk\inst{1}\fnsep\thanks{\email{grzegorz.wilk@ncbj.gov.pl}} \and
        Zbigniew W\l odarczyk\inst{2}\fnsep\thanks{\email{zbigniew.wlodarczyk@ujk.kielce.pl}}
}

\institute{National Centre for Nuclear Research, Ho\.za 69, 00-681 Warsaw, Poland
\and
           Institute of Physics, Jan Kochanowski University, \'Swi\c{e}tokrzyska 15, 25-406 Kielce, Poland
          }

\abstract{The very large transverse momenta and large multiplicities available in present LHC experiments on $pp$ collisions allow a much closer look at the corresponding distributions. Some time ago we discussed a possible physical meaning of apparent log-periodic oscillations showing up in $p_T$ distributions (suggesting that the exponent of the observed power-like behavior is complex). In this talk we concentrate on another example of oscillations, this time connected with multiplicity distributions $P(N)$. We argue that some combinations of the experimentally measured values of $P(N)$ (satisfying the recurrence relations used in the description of cascade-stochastic processes in quantum optics) exhibit distinct oscillatory behavior, not observed in the usual Negative Binomial Distributions used to fit data. These oscillations provide yet another example of oscillations seen in counting statistics in many different, apparently very disparate branches of physics further demonstrating the universality of this phenomenon.}
\maketitle
\section{Introduction}
\label{intro}

To get deeper insight into the dynamics of the multiparticle production processes one measures, in addition to single particle distributions, all kinds of correlations between produced secondaries (i.e., many particle distributions). However, we argue that the very large transverse momenta ( $p_T$ ) and large multiplicities ( $N$ ) available in present LHC experiments on $pp$ collisions already allow for such insight on the level of the corresponding single particle distributions. A first example was provided in \cite{Entropy,ChSF} (and references therein) where it was shown that ratios of $data/fit$ for single particle $p_T$ distributions show distinct, log-periodic oscillations in $p_T$. Because they are present in all LHC experiments, at all energies and for different colliding systems (i.e., also in $Pb Pb$ collisions where they become rather strong for central collisions), it would be unreasonable to regard them as an experimental artifact. Taken seriously,  they strongly suggest that the exponent of the observed power-like behavior of the measured $p_T$ distributions is complex \cite{Entropy,ChSF} (the other possibility would be the existence of similar log-periodic oscillations of the scale parameter present in $p_T$ distributions). We shall not pursue this subject here. Instead we concentrate on another example of oscillations, this time connected with the  multiplicity distributions, $P(N)$. We demonstrate that some combinations of the experimentally measured values of $P(N)$, calculated from recurrence relations widely used in the description of the cascade-stochastic processes in quantum optics, exhibit distinct oscillatory behavior not observed in the usual Negative Binomial Distributions used to fit data \cite{JPG}. These oscillations provide therefore one more example of oscillations seen in counting statistics in many different, apparently very disparate, branches of physics, further demonstrating the universality of this phenomenon \cite{JPG}.

\section{How to fit presently available data}
\label{sec-MNBD}

\begin{figure}[h]
\begin{minipage}{16pc}
\includegraphics[width=16pc]{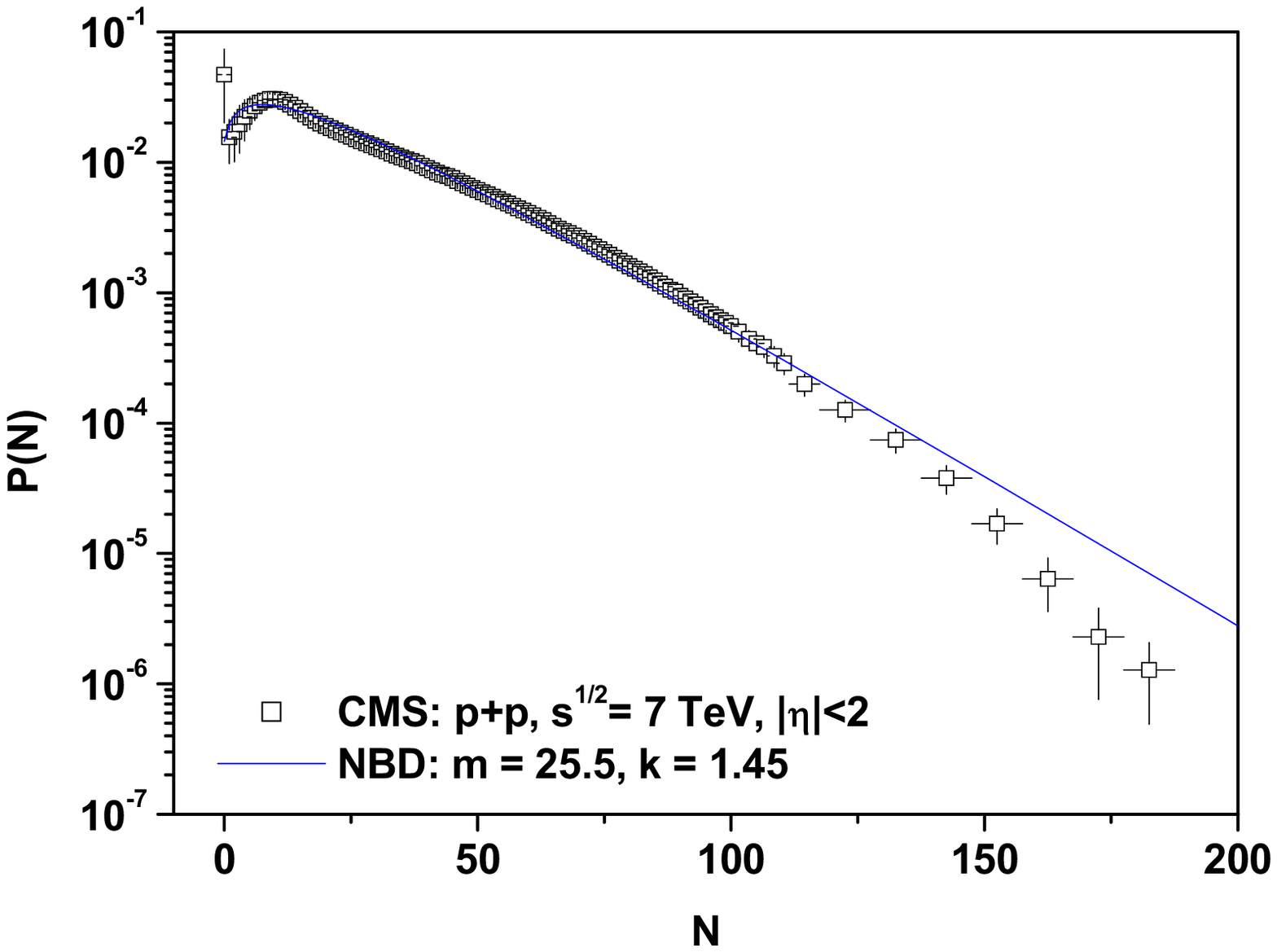}
\vspace{-10mm}
\caption{\label{FF1} (Color online) Charged hadron multiplicity distributions for  $|\eta| < 2$  at $\sqrt{s} =7$ TeV, as given by the CMS experiment \cite{CMS} (points), compared with the NBD for parameters $m = 25.5$ and $k = 1.45$ (solid line).}
\end{minipage}\hspace{1.5pc}%
\begin{minipage}{16pc}
\includegraphics[width=16pc]{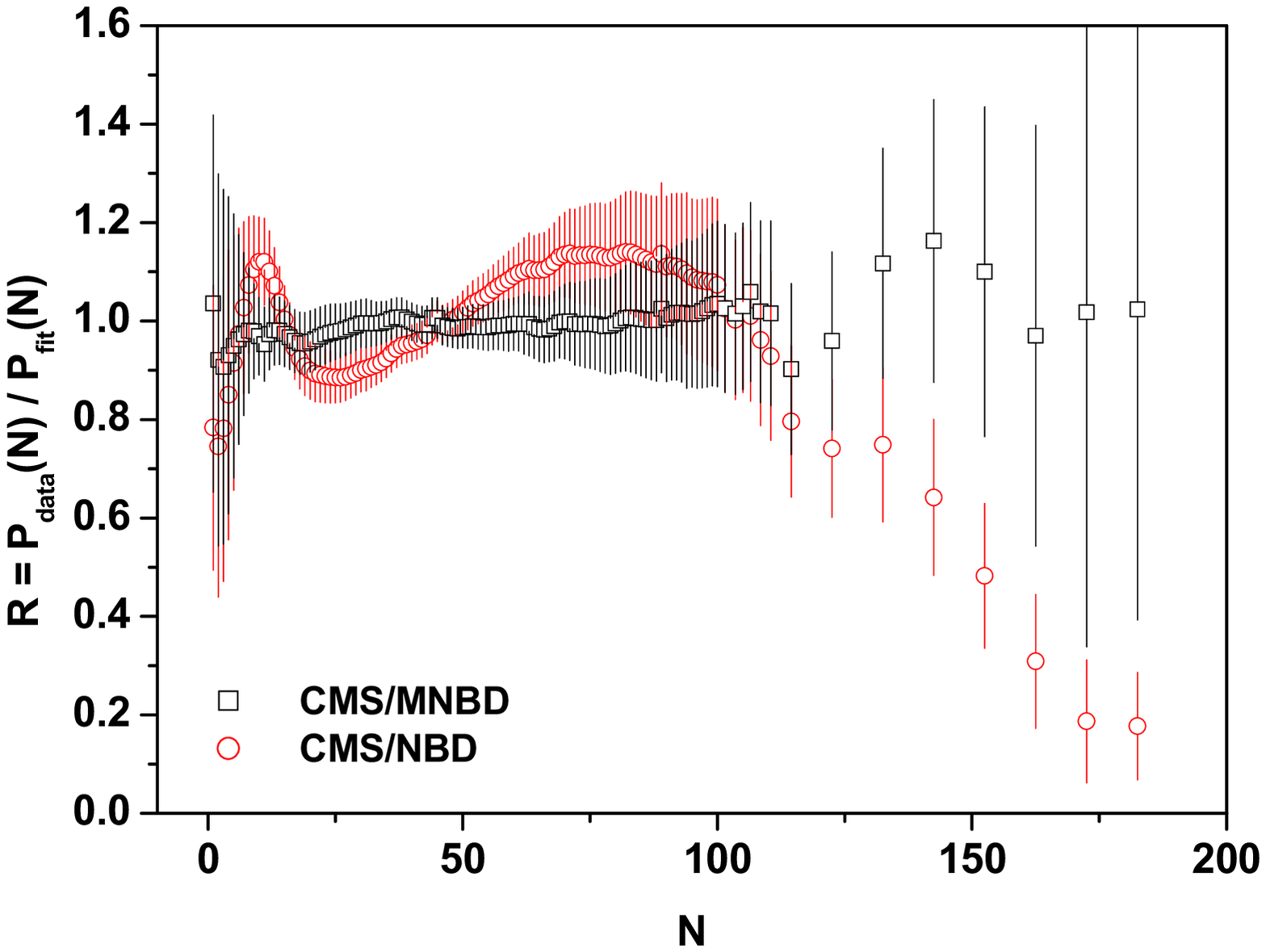}
\vspace{-10mm}
\caption{\label{FF2}(Color online) Multiplicity dependence of the ratio $R = P_{CMS}(N)/P_{NBD}(N)$ for the data shown in Fig. \ref{FF1} (red circles) and of the corrected ratio $R$ (black squares) obtained from the MNBD discussed below.}
\end{minipage}
\end{figure}

Concerning $P(N)$ measured in multiparticle production processes, essentially for all collision energies studied, the most commonly used form is the two-parameter negative binomial distribution (NBD) function (see \cite{Kittel}) and references therein),
\begin{equation}
P(N;p,k) = \frac{\Gamma(N+k)}{\Gamma(k)\Gamma(N+1)}
p^N (1 - p)^k \qquad {\rm with}\qquad  p = p(m,k) = \frac{m}{m + k}, \label{NBD}
\end{equation}
where $p$ is the probability of particle emission, $N$ the observed number of particles and  $m$ and $k$ are the two parameters of the NBD. Whereas $m$ can be connected with the measured multiplicity, $m = \langle N\rangle$ (at least for $m = const$), the actual meaning of $m$ and $k$ depends on the particular dynamical picture behind the NBD form of $P(N)$. However, notwithstanding the popularity of the NBD, one observes systematic (and growing with energy) deviations of fits based on the NBD from data. The example of the recent CMS data \cite{CMS} is shown in Fig. \ref{FF1}; the behavior of the ALICE data \cite{ALICE1} is similar. For large multiplicities the experimental points are below the best single NBD fit. This feature is even more dramatic when one plots the ratio $R = P_{CMS}(N)/P_{NBD}(N)$, see Fig. \ref{FF2} (red circles). In addition to the falling tail one can also see here some structure for smaller multiplicities. Whereas the tails of the distributions in $P(N)$ could be fitted by using weighted (incoherent) superposition of two NBD (cf., for example, \cite{PG,ALICE1} and references therein), the structure for small multiplicities visible in Fig. \ref{FF2} remained to be accounted for.
To get a flat ratio $R$ for the whole measured region of multiplicities $N$, cf., the black squares in Fig. \ref{FF2}, and still adhere to the single NBD form of $P(N)$, one has to modify it. We provide such a modification in \cite{JPG} (and call it the Modified Negative Binomial Distribution - MNBD). It consists of allowing the parameter $m$ in Eq. (\ref{NBD}) to depend on the multiplicity $N$, $m=m(N)$. The desired flat ratio $R=P_{data}(N)/P_{fit}(N)$ can be obtained for
\begin{equation}
m = m(N) = c\exp(a|N-b|) \quad {\rm corresponding~to}\quad p(N) = \frac{1}{1 + \frac{c}{k}\exp ( a |N - b|)} \label{MNBD}
\end{equation}
where $a$, $b$ and $c$ are parameters. Both $m(N)$ and the probability of particle emission $p(N)$ are now non-monotonic functions of $N$ ($p(N)$ = const in the standard NBD). This non-monoticity is located in the region of small multiplicities $N$ \cite{JPG}. Note that such a spout-like form of the proposed modification is just one of the simplest possible choices of parametrization bringing agreement with data. Whereas we cannot at this moment offer any plausible interpretation of such choice, it obviously violates the usual infinite divisibility property of the NBD, bringing it near to the predictions of QCD \cite{H388}.

 \section{Another look at multiplicity distributions $P(N)$}
 \label{Oscillations}

To go further we note that the form of $P(N)$ can be specified by providing the relevant recurrence relation. The simplest one connects only the adjacent distributions, $P(N)$ and $P(N+1)$ (assuming that only neighboring multiplicities can influence each other):
\begin{equation}
(N+1)P(N+1) = g(N)P(N). \label{RR}
\end{equation}
The function $g(N)$ determines the form of the $P(N)$. Its simplest nontrivial linear form, $g(N) = \alpha + \beta N $, covers, for example, the Poissonian distribution (for $ \alpha =  \langle N\rangle$ and $\beta =0$), the binomial distribution (for $\alpha = \langle N\rangle k/(k -\langle N\rangle)$ and $\beta = - \alpha/k$) and the NBD (for $\alpha = \langle N\rangle k/(k +\langle N\rangle)$ and $ \beta = \alpha/k$). When searching for the best $P(N)$  to fit data, $g(N)$ may be modified accordingly (for example, by introducing higher order terms \cite{HC} or by using its more involved forms \cite{MF}), in our MNBD $\alpha = k/[ 1 + k\exp(-a|N-b|)/c]$). However, there also exists a more general form of such a relation, widely used in all processes involving counting statistics (mainly in quantum optics and in \emph{cascade-stochastic processes}), which connects the multiplicity $N+1$ with all smaller multiplicities \cite{ST}:
\begin{equation}
(N +1)P(N + 1) = \langle N\rangle \sum_{j=0}^NC_j P(N-j). \label{recPN}
\end{equation}
Here $P(N)$ is defined by the coefficients $C_j$ (used in the multiparticle phenomenology in \cite{CSF,CSF1}). This recurrence relation provides more information than Eq. (\ref{RR}) and can be regarded as an expansion of $g(N)$ in series of $P(N-j)$ with coefficients $C_j$, which are independent of $N$. Dependence on $N$ is in the $P(N-j)$ themselves and in the limits of summation. For the NBD case $g(N)$ from Eq. (\ref{RR}) is given by
\begin{equation}
g(N) = \frac{\langle N\rangle}{P(N)}\sum_{j=0}^{N} C_j P(N-j) = \frac{mk}{m+k} \frac{\Gamma(N+1)}{\Gamma(N+k)}\sum_{j=0}^{N}\frac{\Gamma(N+k-j)}{\Gamma(N+1-j)} = \frac{m}{m+k}(k+N) \label{g(N)}
\end{equation}
(reproducing the form mentioned before). The important feature of Eq. (\ref{recPN}) is that it can be easily inverted, i.e., knowing (for example, from the measurement process) all $P(N)$, one can use this \emph{experimental information} to deduce the corresponding $C_j$ by means of the following recurrence relation \cite{JPG},
\begin{equation}
C_j = \frac{(j + 1)}{\langle N\rangle} \left[\frac{P(j+1)}{P(0)}\right] - \sum^{j-1}_{i=0}C_i \left[ \frac{P(j-i)}{P(0)}\right]. \label{Cj}
\end{equation}
\begin{figure}[h]
\begin{minipage}{16pc}
\includegraphics[width=16pc]{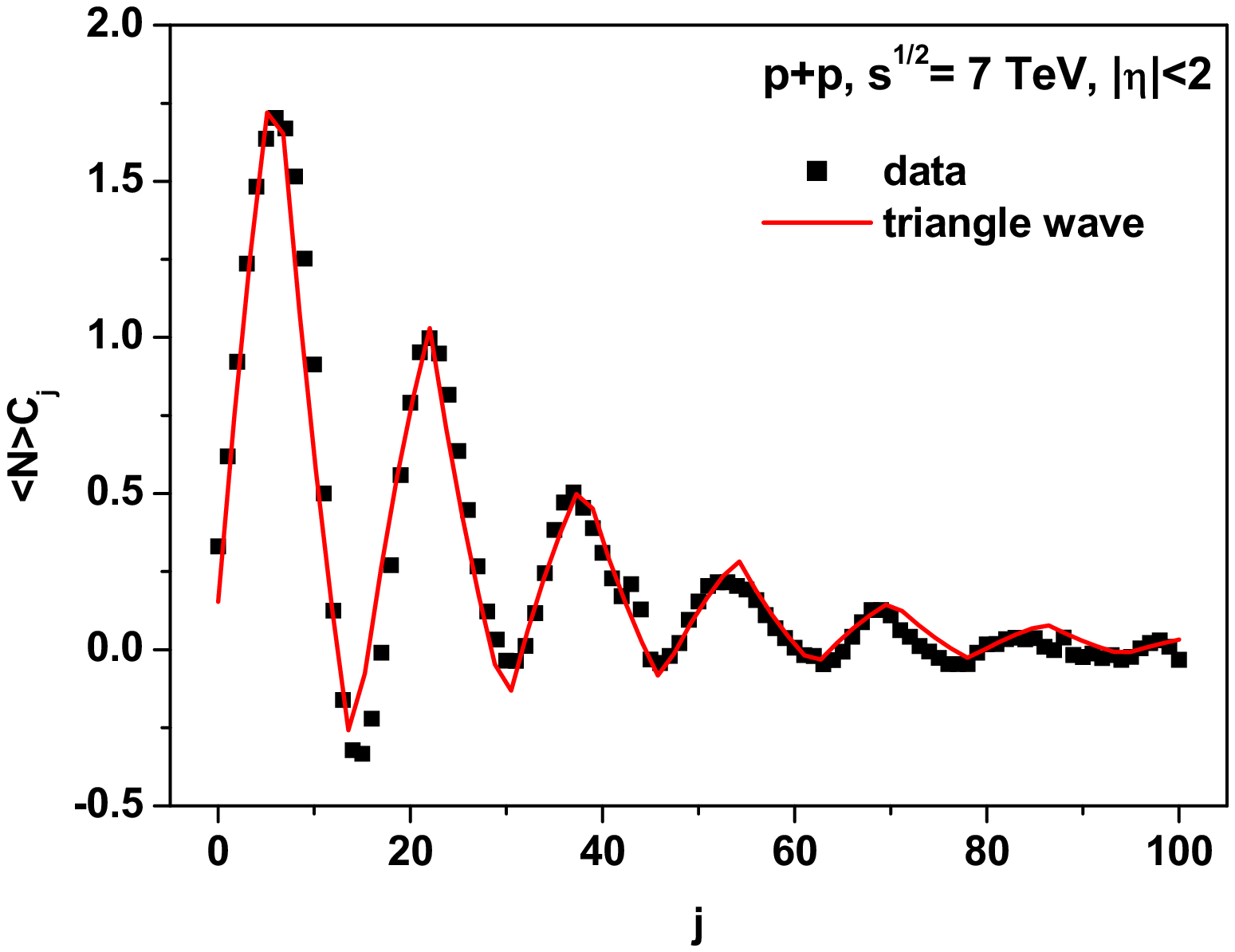}
\vspace{-5mm}
\caption{\label{FF5} (Color online) coefficients $C_j$ obtained from the $pp$ collisions CMS data for $\sqrt{s} = 7$ TeV and pseudorapidity window $|\eta| < 2$ \cite{CMS} fitted using the parametrization given by Eq. (\ref{data}) with parameters $a_1 = 3.2$, $a_2 = 0.6$, $\omega = 16$, $\delta = 1.67$ and $\lambda = 25$.}
\end{minipage}\hspace{1.5pc}%
\begin{minipage}{16pc}
\includegraphics[width=16pc]{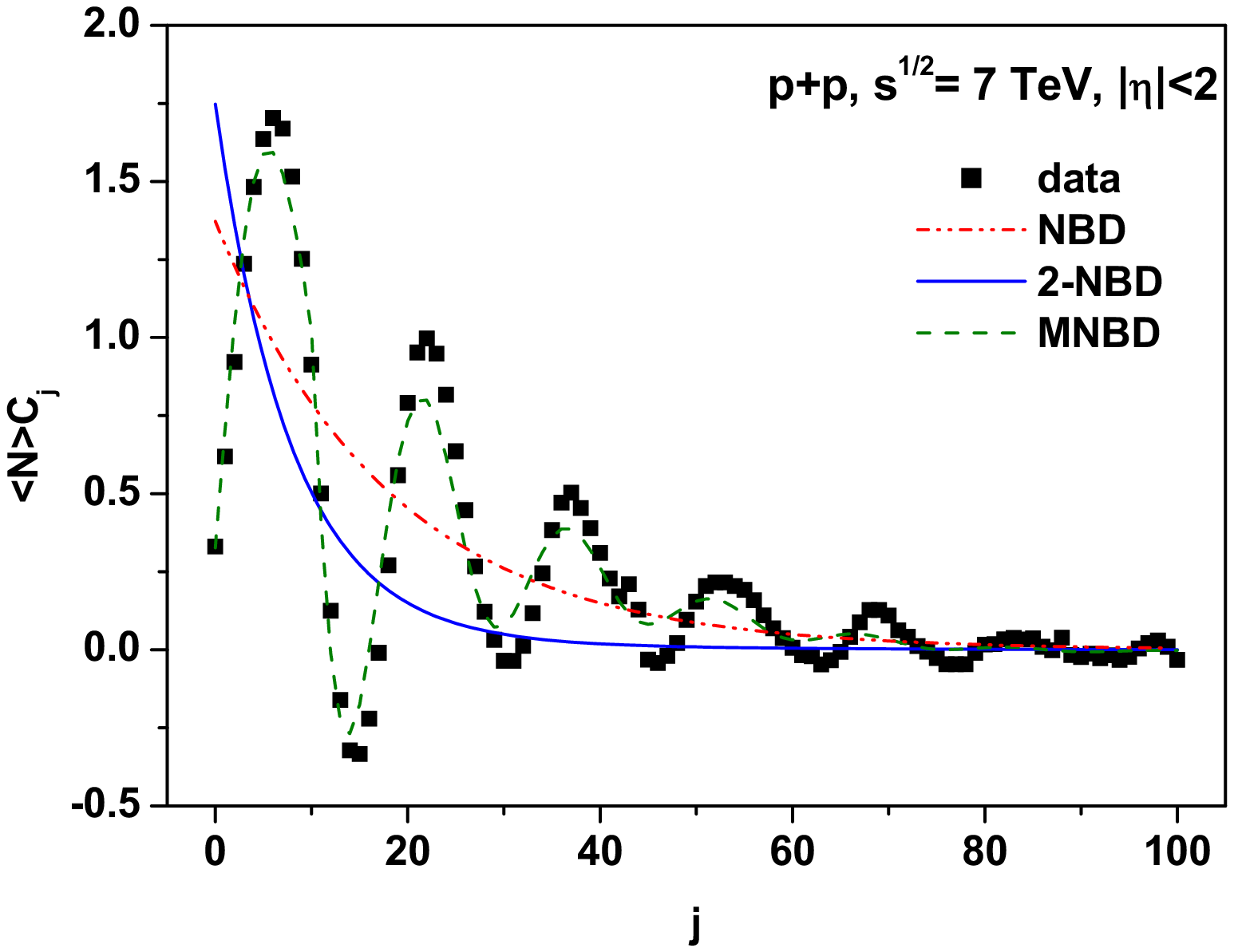}
\vspace{-5mm}
\caption{\label{FF9}(Color online) comparison of the previous $C_j$ with coefficients obtained from different fits to data on $P(N)$: using the single NBD, the $2$-component NBD ($2$-NBD) with parameters from \cite{PG} and our MNBD fit. Only the last choice fits $C_j$.}
\end{minipage}
\end{figure}
In this way one obtains \emph{additional information} on $P(N)$, namely not only their values but also how much they are influenced by the neighboring $P(N-j)$, where $j\in [1,N-1]$. Fig. \ref{FF5} shows clearly that this is, indeed, the case. One observes a distinct oscillatory, exponentially damped, behavior of the $C_j$ obtained from the CMS data \cite{CMS}. This can be fitted by an exponentially damped triangular wave:
\begin{equation}
C_j = \frac{1}{\langle N\rangle}\left\{ a_1 \left[ 1   - \Bigg| 1 -
2\left( \frac{j + \delta}{\omega}- Int\left( \frac{j + \delta}{\omega}
\right)\right)\Bigg| \right] - a_2 \right\}\cdot \exp \left( - \frac{j + \delta}{\lambda}
\right). \label{data}
\end{equation}
(the parameter $\omega$ describes the observed periodicity period). Fig. \ref{FF9} shows the same $C_j$ compared with coefficients obtained from different fits to data on the $P(N)$, starting from the single NBD, using the $2$-component NBD ($2$-NBD) with parameters taken from a fit to the data presented in \cite{PG} and, finally, using our MNBD fit. One can see that only our MNBD fit reproduces the experimentally obtained coefficients $C_j$. In \cite{JPG} it is also shown that the amplitude of these oscillations decrease with the narrowing of the pseudo-rapidity window, $|\eta|$, in which the data are collected (eventually they vanish for  small $|\eta|$) and that the same behavior is also observed for the ALICE data \cite{ALICE1} (albeit with stronger oscillations in this case).
\begin{figure}[h]
\begin{minipage}{16pc}
\includegraphics[width=16pc]{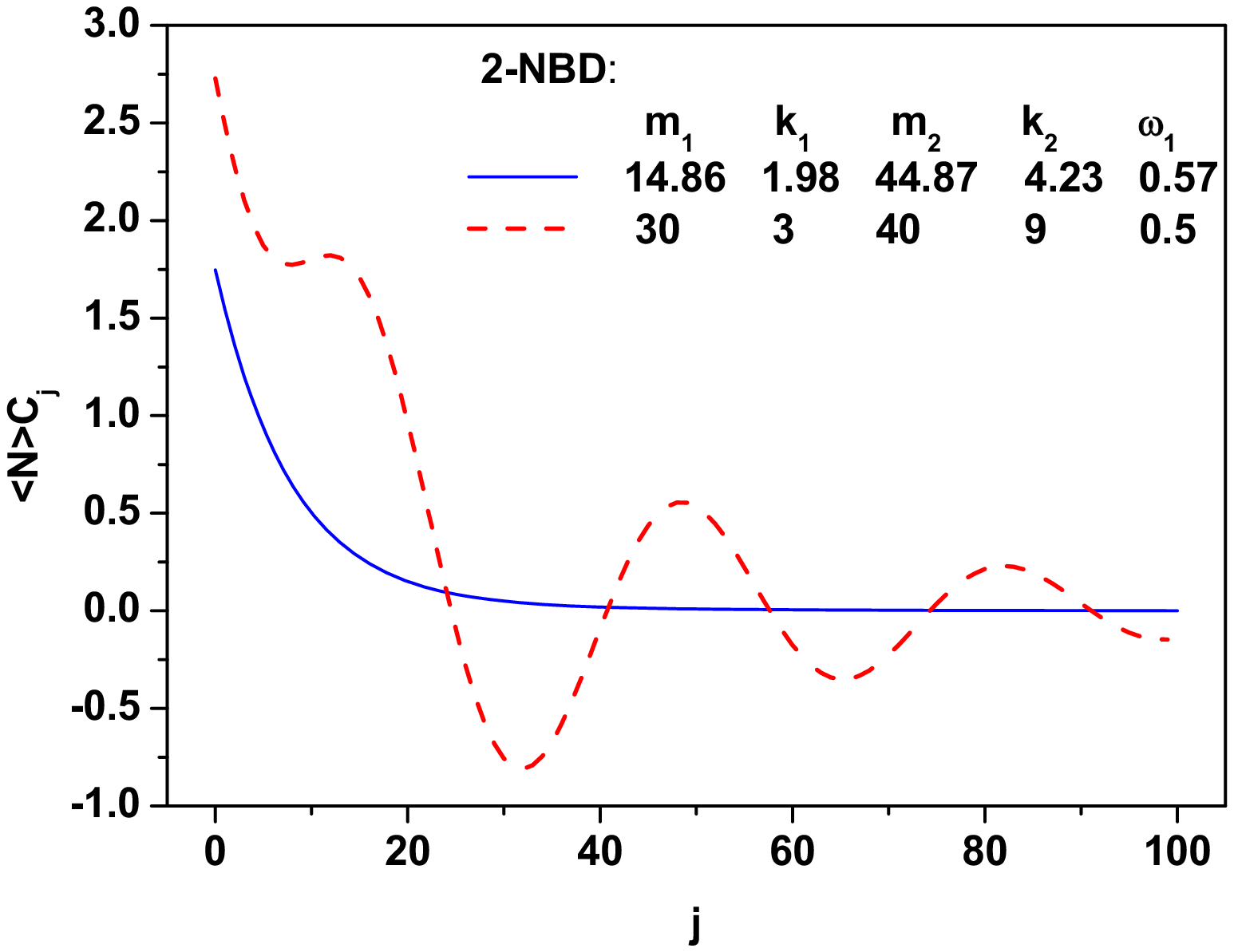}
\vspace{-5mm}
\caption{\label{FF11} (Color online) Coefficients $C_j$ from the $2$-component NBD with parameters from \cite{PG} compared with $C_j$ from the $2-NBD$ with special choice of parameters leading to the oscillations (dashed line). }
\end{minipage}\hspace{1.5pc}%
\begin{minipage}{16pc}
\includegraphics[width=16pc]{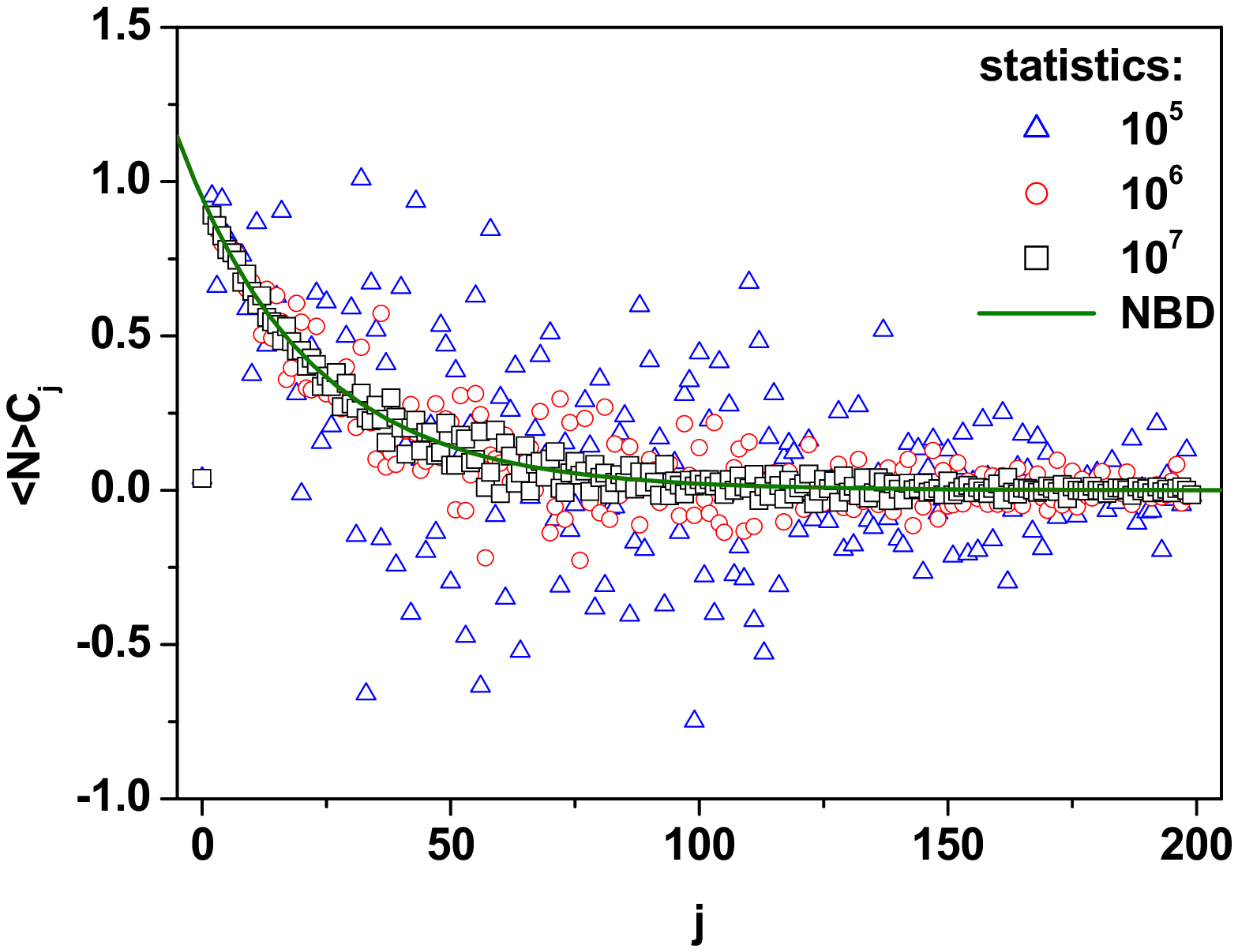}
\vspace{-5mm}
\caption{\label{stat}(Color online) Illustration of dependence on the statistics used. $C_j$ are calculated for $P(N)$ which were constructed using different number of events  sampled from the NBD.}
\end{minipage}
\end{figure}

A word of caution is in order at this point. In the analysis presented so far the measured $P(N)$ are assumed to be fully reliable. Any sensitivity of $C_j$  to the systematic uncertainties of the measurement, to the unfolding uncertainties or, in general, to any other details of the experimental procedure used to obtain $P(N)$, which were subsequently published and which we have used above, can be checked only on the level of experiment, namely by the careful analysis of the raw data, using the proper response matrix, and so on. This exceeds our capability. This means that, in what follows, we feel free to proceed further with our reasoning. Firstly, let us observe in Fig. \ref{FF9} that the coefficients $C_j$ calculated from single or double NBD $P(N)$ used so far to fit the CMS data simply drop monotonically without any sign of fluctuations. However, when we use probabilities of particle emission $p=p(N)$ from MNBD, Eq. (\ref{MNBD}),  with parameters chosen to fit the CMS data, then the coefficients $C_j$ follow \emph{exactly} the oscillatory behavior of the $C_j$ obtained directly from these data, cf. Fig. \ref{FF9}. The observed monotonic behavior of $C_j$ for a single NBD is expected, because in this case
\begin{equation}
C_j = \frac{k}{\langle N\rangle} p^{j+1} = \frac{k}{k+m}\exp(j\ln p),\label{C_j_all}
\end{equation}
and they depend only on the rank $j$ and on the probability of particle emission $p$, which in this case does not depend on $N$ and is smaller than unity. Such $C_j$ reproduce only the monotonically falling damping exponent in Eq. (\ref{data}), with $\lambda = - 1/\ln p$). (It is worth noting that in the case of a binomial distribution (BD) one has $C_j = (-1)^j(k/m)[m/(k - m)]^{(j+1)}$, which oscillate very rapidly with period $2$.). Also in the case of $C_j$ obtained from the $2$-NBD used to fit the CMS data \cite{CMS} (with parameters given in  \cite{PG}) no oscillations occur. However, as shown in Fig. \ref{FF11}, in this case one can find such a set of parameters that the combination of these $2$ NBDs  provides oscillatory behavior for the corresponding parameters $C_j$ (although it does not necessarily fit data on $P(N)$, at least not in the case presented here, which is aimed only to demonstrate the possibility of oscillatory behavior in the case of multi-NBD compositions). As shown in \cite{JPG}, when we have a superposition of a number of NBD, $P(N) = \sum_i \omega_i P_{NBD}\left(N,p_i\right)$, with weights $\omega_i$ (with $\sum_i \omega_i =1$) and emission probabilities $p_i = m_i/\left(m_i + k_i\right)$, then the corresponding coefficients are
\begin{eqnarray}
C_j &=& \frac{1}{P(0)} \sum_i \omega_i p_i^j \left(1 - p_i\right)^{k_i +1}\cdot \nonumber\\
&\cdot& \left\{ \frac{\Gamma\left(j+k_i+1\right)}{\Gamma\left( k_i\! +\! 1\right)\Gamma(j+1)}\frac{m_i\! -\! \langle N\rangle}{\langle N\rangle}\! +\! 1\! +\! \sum_{l=0}^{j-1}\left[ 1\! -\! \frac{C_l}{p_i^l\left( 1 - p_i\right)}\right] \frac{\Gamma\left(j\! -\! l\! +\! k_i\right)}{\Gamma\left(k_i\right)\Gamma\left(j - l +1\right)} \right\}. \label{MNBD2}
\end{eqnarray}
For $m_i < \langle N\rangle$, as well as for $C_l > p_i^l\left(1-p_i\right)$, we have negative terms which can result in nonmonotonic behavior of the coefficients $C_j$.

\section{Summary}
\label{Summary}

To summarize: our results show that a successful model of multiparticle production should describe, with the same set of parameters, both the multiplicity distributions, $P(N)$, and the corresponding coefficients, $C_j$. This is because these coefficients disclose interrelations between $P(N)$ and all $P(N-j)$ with $j<N$, connected with some delicate, intrinsic correlations between multiplicities, which, in turn, depend on the details of the dynamics of multiparticle production. In fact, a similar conjecture was already reached some time ago using the so called {\it combinants}, $C_j^{\star}$ \cite{KG,VVP,BS,H318,H388,H463}. It turns out that our coefficients $C_j$ are directly connected to the combinants, namely
\begin{equation}
C_j = \frac{(j+1)}{\langle N\rangle}C^{\star}_{j+1}. \label{CC}
\end{equation}
Eq. (\ref{CC}) shows that the $C_i$ are much more sensitive to oscillations than the $C_j^{\star}$, especially for higher values of $j$. Note also that our analysis is not directly connected with the wave structure observed in the data on $P(N)$ for multiplicities above $N=25$ \cite{ALICE1}. The coefficients $C_j$ are completely insensitive to the $P(N > j+1)$ tail of the multiplicity distribution, while the oscillatory behavior of $C_j$ is observed starting from the very beginning.

As shown in Fig. \ref{stat} the coefficients $C_j$ can be measured only for events with high enough statistics. That is why they were practically not visible in previous experiments, such high statistics is available only starting from recent LHC experiments. It is also worth pointing out that, in general, $\sum_{j=1}^{\infty} C_j \ge 1$ (this is direct consequences of the fact that $\sum_{N=j}^{\infty} P(N-j) \leq 1$). The strict equality holds in the case of $P(N)$ in the form of NBD. Note that in the case of combinants $C_j^{\star}$ one has that for the NBD $\sum_{j=1}^{\infty} C_j^{\star} = -k \ln [k/(k+m)]$.

Finally, let us note the experimental drawback of the $C_j$. This is the necessity to know (measure) the $P(0)$, cf. Eq. (\ref{Cj}), which serves as a normalization factor in our approach. This is not an easy (to say the least) point in any experiment because $P(0)$ depends critically on the acceptance (cf. the discussion in \cite{JPG}).

\medskip
Acknowledgements: We are indebted to Sandor Hegyi, Edward Grinbaum-Sarkisyan and Adam Jacho\l kowski for fruitful discussions and we would like to thank warmly Nicholas Keeley for reading the manuscript.


\begin{thebibliography}{}

\bibitem{Entropy} G.Wilk and Z.W\l odarczyk, Entropy \textbf{17}, 384 (2015).

\bibitem{ChSF} G.Wilk and Z.W\l odarczyk, Chaos, Solitons and Fractals {\bf 81}, 487 (2015).

\bibitem{JPG} G.Wilk and Z.W\l odarczyk, \textit{How to retrieve additional information from the multiplicity distributions},
arXiv:1601.03883.

\bibitem{Kittel} W.Kittel and E.A.De Wolf, \textit{Soft Multihadron Dynamics}, World Scientific, Singapore, 2005.

\bibitem{CMS} V.Khachatryan {\it et al.} (CMS Collaboration), J. High Energy Phys.\textbf{01}, 079 (2011).

\bibitem{ALICE1} J.Adam {\it et al.} (ALICE Collaboration), \textit{Charged particle multiplicities
                 in proton-proton collisions at $\sqrt{s} = 0.9$ to $8$ TeV}, arXiv:1509.07541.

\bibitem{PG} P.Ghosh, Phys. Rev. D \textbf{85}, 0541017 (2012).

\bibitem{H388} S.Hegyi, Phys. Lett. B {\bf 388}, 837 (1996).

\bibitem{HC} T.F.Hoang and B.Cork, Z. Phys. C {\bf 36}, 323 (1987).

\bibitem{MF} S.V.Chekanov and V.I.Kuvshinow, J. Phys. G {\bf 22}, 601 (1996).

\bibitem{ST} B.E.A.Saleh and M.K.Teich, Proc. IEEE {\bf 70}, 229 (1982).

\bibitem{CSF} V.D.Rusov, T.N.Zelentsova, S.I.Kosenko, M.M.Ovsyanko and I.V.Sharf, Phys. Lett. B {\bf 504}, 213 (2001).

\bibitem{CSF1} V.D.Rusov and I.V.Sharf, Nucl. Phys. A {\bf 764} 460 (2006).

\bibitem{KG} S.K.Kauffmann and M.Gyulassy, J. Phys. A {\bf 11}, 1715 (1978).

\bibitem{VVP} R.Vasudevan, P.R.Vittal and K.V.Parthasarathy, J. Phys. A {\bf 17} 989 (1984).

\bibitem{BS} A.B.Balantekin and J.E.Seger, Phys. Lett. B {\bf 266} 231 (1991).

\bibitem{H318} S.Hegyi, Phys. Lett. B {\bf 318}, 642 (1993).

\bibitem{H463} S.Hegyi, Phys. Lett. B {\bf 463}, 126 (1999).

\end{thebibliography}
\end{document}